\documentclass[]{jd04}    
\usepackage{natbib}
\usepackage{graphicx}

\Title[White Dwarfs in GALEX Survey]{White Dwarfs in the GALEX Survey}

\Author{Kawka$^1$, Adela}
\Author{Vennes$^2$, St\'ephane}

\Address{$^1$Astronomick\'y \'ustav AV {\v C}R, CZ-25165 Ond{\v r}ejov, Czech Republic \\
$^2$Department of Physics and Space Sciences, Florida Institute of Technology, Melbourne, Florida 32901-6975, USA \\
}{}

\ShortAuthors{Kawka \& Vennes}


\begin{document}
\maketitle

\begin{abstract}
We have cross-correlated the 2dF QSO Redshift Survey (2QZ) white dwarf
catalog with the GALEX 2nd Data Release and the Sloan Digital Sky Survey (SDSS)
data release 5 to obtain ultraviolet photometry (FUV, NUV) for approximately
700 objects and optical photometry ({\it ugriz}) for approximately 800 
objects. We have compared the optical-ultraviolet colors to synthetic white 
dwarf colors to obtain temperature estimates for approximately 250 of these 
objects. These white dwarfs have effective temperatures ranging from 10\,000 K 
(cooling age of about 1Gyr) up to about 40\,000 K (cooling age of about 3 Myrs),
with a few that have even higher temperatures. We found that to distinguish 
white dwarfs from other stellar luminosity classes both optical and ultraviolet 
colors are necessary, in particular for the hotter objects where there is 
contamination from B and O main-sequence stars. Using this sample we build a 
luminosity function for the DA white dwarfs with $M_V < 12$ mag.
\end{abstract}

\section{Introduction}

The properties of white dwarf stars are relatively simple to measure.
Ideally, a complete spectral energy distribution, from the optical to the
ultraviolet, helps establish the white dwarf atmospheric properties, i.e.,
their effective temperature, surface gravity, and chemical composition,
from which we may deduce their age and mass using theoretical mass-radius
relations \citep[e.g.,][]{woo1995}. These measurements lay the foundations for 
a study of the white dwarf cooling history (luminosity function).

The 2dF QSO Redshift Survey (2QZ) survey is a deep spectroscopic survey of blue
Galactic and extra-Galactic sources which was conducted at the Anglo-Australian
Telescope (AAT). Over 2000 white dwarfs were discovered during this survey
\citep{cro2004}. \citet{ven2002} obtained atmospheric parameters for many of 
these objects. Another survey is the Sloan Digital Sky Survey (SDSS),
which aims to observe 1/4 of the sky in {\it ugriz} photometric bands and
obtain spectroscopy for many of these objects. A catalog of 9316
spectroscopically confirmed white dwarfs from the SDSS 4th data release
was published by \citet{eis2006}.
The Galaxy Evolution Explorer (GALEX) is an orbiting observatory that aims to
observe the sky in the ultraviolet. GALEX provides photometry in two bands,
{\it FUV} (1344-1786 \AA) and {\it NUV} (1771-2831 \AA), and will conduct 
several surveys during its mission, including the
All-Sky Imaging Survey (AIS) with a limiting magnitude of 21.5 mag and the
Medium-Imaging Survey (MIS) with a limiting magnitude of 23 mag.

\begin{figure}
\centering
\includegraphics[width=0.6\textwidth]{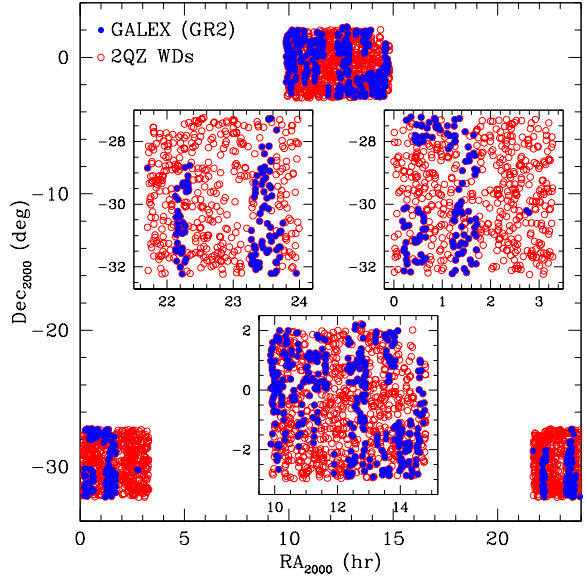}
\caption{The overlap between 2QZ and GALEX (GR2) surveys.\label{fig_coord_comp}}
\end{figure}

We cross-correlated the 2QZ DA white dwarf catalog with the SDSS 5th
data release to obtain {\it ugriz} photometry. Only the north Galactic cap
(NGP) is covered by the SDSS and therefore we were only able to get {\it ugriz}
photometry for 795 objects. We also cross-correlated the 2QZ DA white dwarf
catalog with the GALEX 2nd data release (GR2) to obtain ultraviolet photometry
for approximately 810 objects, however some of these stars have only been
detected in either the FUV or NUV band. Figure~\ref{fig_coord_comp} shows 
the overlap between the 2QZ white dwarf catalog and GR2. Finally we 
cross-correlated our SDSS and GALEX samples to obtain 252 stars for which we 
have both optical {\it ugriz} and ultraviolet photometry.

\section{Determining the parameters}

\begin{figure}
\includegraphics[width=0.5\textwidth]{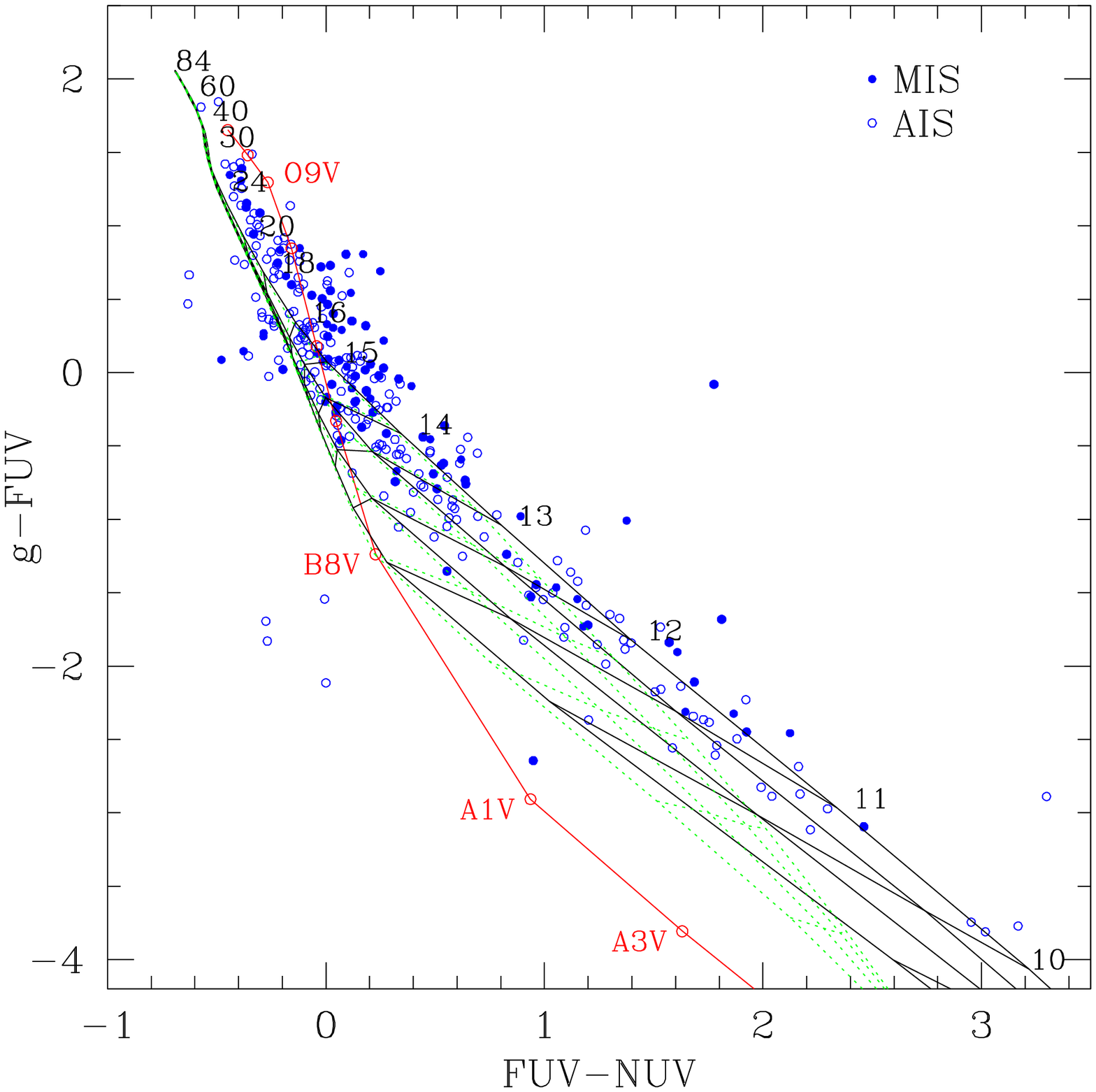}
\includegraphics[width=0.5\textwidth]{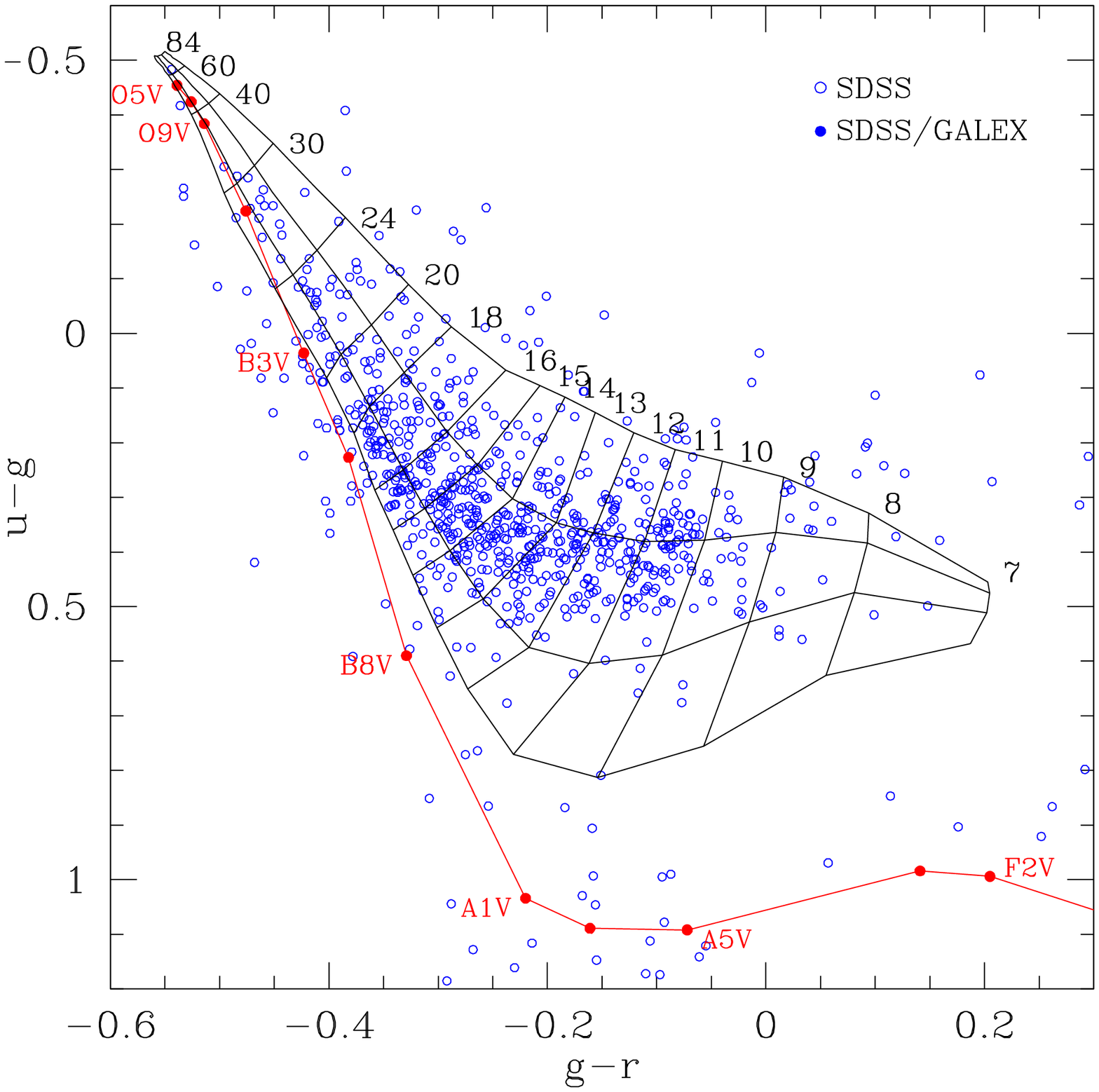}
\caption{({\it left}) Optical and ultraviolet colors ($g-FUV$ vs. $FUV-NUV$) of 2QZ
white dwarfs compared to synthetic DA white dwarf colors. The full-lined grid 
includes the effect of Ly$\alpha$ satellites, and the dotted-lined grid excludes
them. ({\it right}) Optical ($u-g$/$g-r$) colors of 2QZ white dwarfs compared
to synthetic DA white dwarf colors. The effective temperature is indicated in 
units of 1000 K and the $\log{g} = 6.0$, $7.0$, $8.0$, and $9.0$ ({\it from 
bottom to top}). The main-sequence colors 
are also shown.\label{fig_gmf_fmn}}
\end{figure}

We obtained effective temperatures for the 2QZ DA white dwarfs by comparing
the ultraviolet/optical colors ($g-FUV$/$FUV-NUV$) and optical colors
($u-g$/$g-r$) to synthetic DA colors. Figure~\ref{fig_gmf_fmn} shows the
ultraviolet and optical colors of the 2QZ white dwarfs compared to synthetic
DA and main-sequence colors. The optical and ultraviolet photometry of the 2QZ 
white dwarfs were corrected for interstellar 
extinction using the dust maps of \citet{sch1998} and the extinction
law of \citet{car1989}, where we have assumed $R_V= 3.1$. Note that $R_V$
can vary anywhere between $\sim 2.5$ to $\sim 5.5$ which can change the
extinction in the ultraviolet by a very significant amount \citep{car1989}.
In addition, the true extinction toward a star is possibly only a fraction
of the total extinction in the line of sight.
The synthetic DA colors were calculated using a grid of pure hydrogen LTE
models \citep{kaw2006}. The grid of models extend from $T_{\rm eff} = 7000$ to 
84\,000 K at $\log{g} =6.0$ to $9.5$. 
The spectra include the effect of Ly$\alpha$ satellites \citep{all1992},
however we also calculated spectra which exclude the effect. 
Figure~\ref{fig_gmf_fmn} (left) shows the effect of Lyman satellites on UV 
colors, which is most prominent in the cooler white dwarfs 
($T_{\rm eff} \lesssim 12000$ K). The color diagrams also show the
main-sequence, which was calculated using Kurucz synthetic spectra
\citep{kur1993}. Figure~\ref{fig_gmf_fmn} (right) shows some stars that have 
colors corresponding to the main-sequence rather than the DA white dwarf 
sequence, therefore some contamination by main-sequence stars is still
present in the 2QZ DA catalog. We found that the combination of both diagrams
helps distinguish between white dwarfs and main-sequence stars.

\begin{figure}
\includegraphics[viewport=0 285 570 570, width=\textwidth]{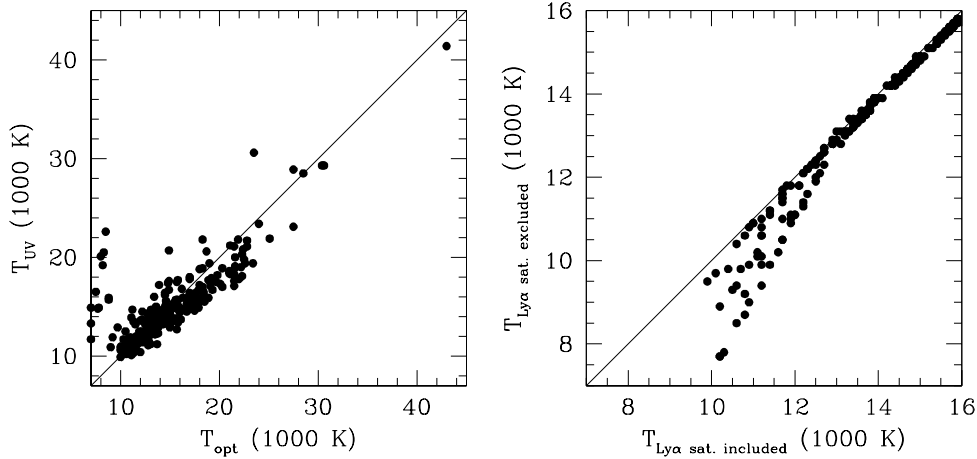}
\caption{In the {\it left} panel, the temperatures obtained from the
optical-ultraviolet and optical diagrams are compared.
And in the {\it right} panel, the temperatures determined using the
optical-UV colors determined if Ly$\alpha$ satellites are included compared
to the temperatures determined using the same grid but which excludes the
effect of the Ly$\alpha$ satellites.\label{fig_temp_lysat}}
\end{figure}

Figure~\ref{fig_temp_lysat} shows a comparison between the
temperatures obtained using the $g-FUV$/$FUV-NUV$ colors and the $u-g$/$g-r$
colors. A relatively good agreement between the temperatures is observed
with a few points which appear to give very low optical temperatures compared
to their UV temperatures. A check of the 2QZ spectra for these objects
reveal them to be DA white dwarfs with cool companions and therefore the UV
temperatures should be adopted as the white dwarf temperatures. 
Figure~\ref{fig_temp_lysat} also shows a comparison of the UV temperatures 
determined with grids which include or exclude the effect of Ly$\alpha$ 
satellites. Temperatures determined using models which exclude the effect of 
Ly$\alpha$ satellites are significantly lower than those which do not for 
$T_{\rm eff} \lesssim 12000$ K.

\section{Luminosity Function}

\begin{figure}
\centering
\includegraphics[width=0.65\textwidth]{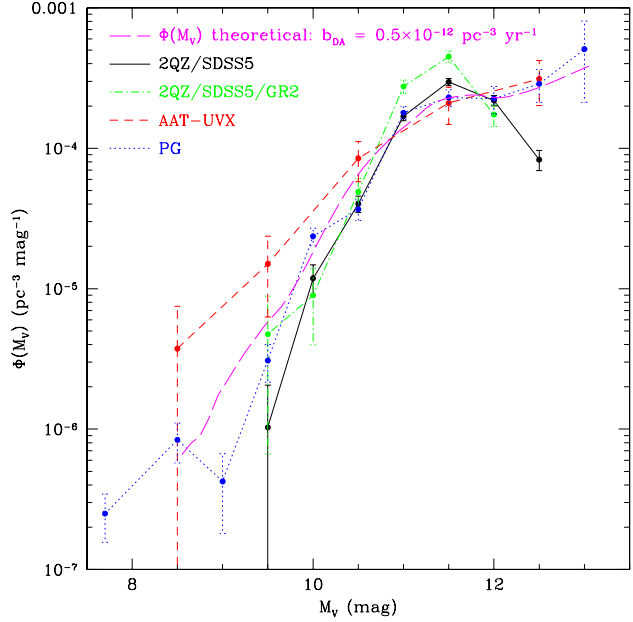}
\caption{The DA luminosity function measured using the 2QZ/SDSS/GR2
survey compared to the luminosity function measured using AAT-UVX and
Palomar-Green (PG) survey. The theoretical luminosity function assuming
a DA white dwarf birthrate of $0.5\times 10^{-12}$ pc$^{-3}$ yr$^{-1}$.\label{fig_lum_GR2}}
\end{figure}

We constructed the DA luminosity function using the
2QZ/SDSS5/GR2 and the 2QZ/SDSS5 samples by using
the accessible-volume method \citep[see][and references therein]{boy1989}
and assuming a scale-height of 250 pc. Using the temperatures obtained from
the $g-FUV$/$FUV-NUV$ and $u-g/g-r$ diagrams, we calculated the absolute 
magnitudes ($M_V$), assuming $\log{g} =8.0$ and using the mass-radius 
relations of \citet{ben1999}. Figure~\ref{fig_lum_GR2} shows the 2QZ/SDSS5 
and 2QZ/SDSS5/GR2 luminosity functions compared to the luminosity 
functions determined in the PG Survey \citep{fle1986}, the AAT-UVX survey 
\citep{boy1989} and a theoretical luminosity function based in the cooling 
sequence of \citet{woo1995} and a constant DA birthrate of 
$0.5\times 10^{-12}$ pc$^{-3}$ yr$^{-1}$. Both the 2QZ/SDSS5/GR2 and
2QZ/SDSS5 luminosity functions are incomplete at the cool end
($M_V \ge 12.0$ mag). The two main reasons for this is that fewer objects
were selected due to the limiting colors at the cool end and the cooler
objects are very faint in the ultraviolet and hence would not be detected
by the GALEX survey. Also there appears to be an excess of stars in the
$M_V = 11.5$ bin, in particular in the 2QZ/SDSS5/GR2 sample. Interstellar
reddening could be a contributing toward this excess since our extinction 
correction did not consider the distance toward the star, and therefore
cooler objects which would be closer would be over corrected resulting in
higher temperatures.

\section{Summary and Future Work}

We have presented our initial analysis of DA white dwarfs from the 2QZ survey 
using SDSS5 and GR2 photometry. Using these photometric data, we obtained
temperature and absolute magnitudes from which we built a luminosity function.
We will extend this analysis to the He-rich sequence of white dwarfs (DO/DB). 
We will investigate the effect of interstellar extinction on the
temperature distribution and hence the luminosity function in more detail.
Also, we will examine the effect of heavy elements on the UV/optical
temperature scales.
Many hot white dwarfs show heavy element lines, and the abundance of
these elements can vary many orders of magnitude.

{\it Acknowledgements:}

This research is supported by NASA/GALEX grant NNG05GL42G.
A.K. is supported by grant GA \v{C}R 205/05/P186.

\end{document}